\documentclass[preprint,12pt]{elsarticle}

\usepackage[T1]{fontenc}
\usepackage{amsmath,amssymb,amsthm}
\usepackage{bm}
\usepackage{booktabs}
\usepackage{graphicx}
\usepackage{enumitem}
\usepackage{microtype}
\usepackage{newtxtext,newtxmath}
\usepackage{hyperref}

\hypersetup{
    hidelinks,
    breaklinks=true,
    pdfcreator={Tectonic},
}
\biboptions{sort&compress}

\journal{Journal of Computational Physics}

\title{Bias Inheritance in Neural-Symbolic Discovery of Constitutive Closures Under Function-Class Mismatch}
\newcommand{\codeurl}{\url{https://github.com/ToughClimb/closure-bias-inheritance}}

\newtheorem{proposition}{Proposition}
\newtheorem{remark}{Remark}

\begin{document}

\begin{frontmatter}

\author[aff1]{Hanbing Liang}
\author[aff1]{Ze Tao}
\author[aff1]{Fujun Liu\corref{cor1}}
\ead{fjliu@cust.edu.cn}
\cortext[cor1]{Corresponding author.}
\address[aff1]{Nanophotonics and Biophotonics Key Laboratory of Jilin Province, School of Physics, Changchun University of Science and Technology, Changchun 130022, P.R. China}

\begin{abstract}
We investigate the data-driven discovery of constitutive closures in nonlinear reaction-diffusion systems where the governing partial differential equation (PDE) structure is known. Our objective is to robustly recover diffusion and reaction laws from spatiotemporal observations while mitigating a common pitfall in scientific machine learning where low data-fit residuals or short-horizon predictions are conflated with physically correct recovery. To address this challenge, we propose a three-stage neural-symbolic framework comprising weak-form-driven numerical constitutive recovery, restricted symbolic compression, and forward re-simulation validation. Specifically, the first stage learns numerical surrogates for the unknown closures under physical constraints such as positivity and smoothness using a noise-robust and weak-form-driven hybrid objective. Subsequently, the second stage compresses these learned surrogates into restricted interpretable families including polynomial, rational, and saturation-style forms. The final stage then validates the symbolic closures through explicit forward simulation on unseen initial conditions. Extensive numerical experiments on one-dimensional systems reveal two distinct behavior regimes. Under matched-library settings, weak polynomial baselines behave as correctly specified reference estimators, demonstrating that neural surrogates should not be assumed to uniformly outperform classical bases. Conversely, under function-class mismatch involving model-form error, neural surrogates provide necessary flexibility and can be compressed into compact symbolic laws with minimal additional surrogate error or rollout degradation. However, we identify a critical mechanism of bias inheritance where symbolic compression does not automatically repair constitutive bias. Across clean, noisy, and sparse observation regimes, the true error of the symbolic closure closely tracks that of the neural surrogate, yielding a bias inheritance ratio near one. These findings ultimately demonstrate that the primary bottleneck in neural-symbolic modeling lies in the initial numerical inverse problem rather than the subsequent symbolic compression, underscoring that constitutive claims must be rigorously supported by forward validation rather than residual minimization alone. Code, manuscript source, and selected paper-facing artifacts are available at \codeurl.
\end{abstract}


\end{frontmatter}

\section{Introduction}

Many physical and engineering systems are governed by partial differential equations where the macroscopic conservation laws are well established, yet the underlying constitutive closures remain elusive. This scenario is ubiquitous across transport phenomena, continuum mechanics, and data-driven closure modeling for multiscale and fluid systems \cite{Kirchdoerfer_2016,Kirchdoerfer_2017,Duraisamy_2019,Brunton_2020}. In such contexts, while the governing balance equations dictate the system's structural evolution, the specific diffusion, mobility, or reaction terms must be inferred from observational data. Consequently, the primary objective of data-driven modeling in these domains extends beyond building predictive simulators to discovering the fundamental constitutive laws themselves.

To address this challenge of equation discovery, sparse identification techniques and data-driven methodologies have emerged as central paradigms within scientific machine learning \cite{Bongard_2007,Schmidt_2009,Brunton_2016,Rudy_2017,Schaeffer_2017,Reinbold_2020}. This foundational work has been rapidly expanded by the development of physics-informed neural networks, hidden-physics inverse solvers, and physically constrained optimization frameworks designed to integrate known governing equations with neural model-discovery pipelines \cite{Raissi_2018,Raissi_2019,Kaheman_2020,Champion_2020,Long_2019,Both_2021,Karniadakis_2021}. Within this landscape, symbolic regression has proven particularly attractive for closure modeling, as the ultimate goal is not merely to construct a black-box surrogate but to extract an interpretable, algebraic constitutive relation that can be rigorously analyzed \cite{Udrescu_2020,Cranmer_2020,Biggio_2021,Petersen_2021,LaCava_2021}.

Building upon these neural-symbolic advancements, the present study focuses on the inverse problem of constitutive discovery in nonlinear reaction-diffusion systems. Specifically, we consider governing equations of the form
\begin{equation}
u_t = \partial_x \left( D(u) u_x \right) + R(u),
\label{eq:rd_main}
\end{equation}
where the overarching spatial and temporal derivative structure is known, but the specific nonlinear constitutive relations for diffusion $D(u)$ and reaction $R(u)$ remain entirely unspecified. Because reaction-diffusion systems serve as canonical models for complex pattern formation and morphogenesis \cite{Turing_1952,Murray_1989,Kondo_2010}, they provide a rigorous testing ground for inverse modeling techniques. Given spatiotemporal observations of the state variable $u(x,t)$, our objective is to uniquely recover these unknown closures in a mathematical form that is numerically robust, scientifically interpretable, and verifiable through forward dynamic simulation.

Achieving such verifiable recovery is significantly complicated by the inherent nature of closure discovery, which functions fundamentally as an ill-posed inverse problem rather than a standard predictive machine learning task. Minimizing the residual error on observed trajectories or achieving a small short-horizon rollout error does not mathematically guarantee the correct identification of the underlying physics. Under limited dynamic excitation, different combinations of diffusion and reaction closures may compensate for one another across the observed state space, yielding highly similar trajectory-level predictions despite severe constitutive inaccuracies. This phenomenon creates a critical gap between merely fitting observational data and genuinely identifying the true physical law, a structural challenge that aligns with established literature on limited informativeness and non-uniqueness in inverse problems \cite{Tarantola_2005,Isakov_2006,Stuart_2010}. Bridging this gap between trajectory fitting and true constitutive recovery constitutes the primary methodological focus of this work.

Recognizing this fundamental gap necessitates a shift in how data-driven constitutive models are evaluated and constructed. In matched-library scenarios where the true closure is known to reside within a restricted polynomial dictionary, classical weak-form polynomial baselines act as correctly specified reference estimators and should be treated as rigorous benchmarks rather than naive comparators \cite{Brunton_2016,Rudy_2017}. However, when severe function-class mismatch occurs and the true physical law lies outside these restricted libraries, highly parameterized neural networks become indispensable as flexible numerical approximators. Nevertheless, a learned neural surrogate is inherently opaque and does not constitute a mature scientific law until it is distilled into a compact mathematical expression and subsequently validated. To execute this distillation, we propose a comprehensive three-stage neural-symbolic framework for closure discovery \cite{Long_2019,Cranmer_2020,Udrescu_2020,Biggio_2021}. The initial stage learns stable numerical constitutive surrogates by optimizing a robust, weak-form-driven hybrid objective. The second stage then compresses these learned numerical surrogates into restricted and interpretable symbolic families. Finally, the third stage reinserts these symbolic closures back into the partial differential equation to explicitly validate the extracted laws via forward re-simulation on unseen initial conditions.

By decomposing the discovery process into these explicit stages, this framework allows us to isolate the exact source of constitutive error and systematically evaluate the broader neural-symbolic pipeline. Our theoretical analyses and numerical experiments demonstrate that while symbolic compression successfully preserves learned dynamic behavior with minimal additional approximation error, it does not inherently repair the constitutive bias generated by the neural surrogate. Across a variety of clean, noisy, and sparse observation environments, the true error of the symbolic closure consistently tracks the error of the initial neural surrogate, resulting in a bias inheritance ratio that remains persistently near unity. Consequently, this study yields several distinct contributions to the field of data-driven mechanics. First, we formulate constitutive closure discovery under a known governing structure as a purely weak-form-driven inverse problem, placing explicit emphasis on the constraints of system identifiability, dynamic excitation coverage, and function-class mismatch. Second, we introduce a decoupled neural-symbolic pipeline that strictly separates numerical approximation, symbolic compression, and forward equation validation. Third, we establish a theoretical foundation explaining why low observation residuals do not guarantee true recovery, why polynomial baselines exhibit exceptional strength in correctly specified regimes, and why symbolic regression functionally preserves rather than eliminates numerical bias. Ultimately, by benchmarking this framework against established reference estimators, we conclude that the dominant operational bottleneck in neural-symbolic discovery lies in the initial numerical inverse problem rather than the subsequent compression phase, dictating that closure claims must be rigorously substantiated by explicit forward validation rather than residual minimization alone.

\section{Problem Formulation}

We consider one-dimensional reaction-diffusion systems on the periodic domain $\Omega = [0,1)$:
\begin{equation}
u_t = \partial_x \left( D(u) u_x \right) + R(u),
\qquad x \in \Omega, \quad t \in [0,T].
\end{equation}
The field $u(x,t)$ is observed, while the closure pair
\begin{equation}
f = (D,R)
\end{equation}
is unknown. The task is to recover $D(u)$ and $R(u)$ from spatiotemporal data generated by multiple trajectories with varying initial conditions.

We study this problem under two regimes.
\begin{itemize}[leftmargin=1.5em]
\item \textbf{Matched-library regime.} The true closure lies in a restricted polynomial family. This is the setting of Case~A and Case~B and serves as a correctly specified reference regime.
\item \textbf{Function-class-mismatch regime.} The true closure lies outside the restricted polynomial library. This is the setting of Case~Exp and is the main stress test for the neural-symbolic pipeline.
\end{itemize}

The data consist of finite trajectory sets
\begin{equation}
\mathcal{D} = \left\{ u^{(m)}(x,t_n) \right\}_{m=1}^{M},
\end{equation}
possibly corrupted by Gaussian noise and/or spatial-temporal downsampling. The objective is not only to fit these trajectories, but to recover closures that remain valid under symbolic compression and forward rollout on unseen initial conditions.

We therefore evaluate every method at three levels.
\begin{enumerate}[leftmargin=1.5em]
\item \textbf{Trajectory level:} weak residual and unseen rollout error.
\item \textbf{Closure level:} relative error of $D(u)$ and $R(u)$ on a common support grid.
\item \textbf{Neural-symbolic level:} error between the symbolic closure and the learned neural surrogate, plus the post-compression rollout error.
\end{enumerate}

For the closure-level metrics, we evaluate both true and recovered closures on a common support grid $\{u_\ell\}_{\ell=1}^L \subset \mathcal{U}$ over the relevant state interval and report
\begin{align}
\mathrm{ErrD}
&=
\frac{\left(\sum_{\ell=1}^{L} |D(u_\ell)-\widehat{D}(u_\ell)|^2 \right)^{1/2}}
{\left(\sum_{\ell=1}^{L} |D(u_\ell)|^2 \right)^{1/2} + 10^{-12}}, \\
\mathrm{ErrR}
&=
\frac{\left(\sum_{\ell=1}^{L} |R(u_\ell)-\widehat{R}(u_\ell)|^2 \right)^{1/2}}
{\left(\sum_{\ell=1}^{L} |R(u_\ell)|^2 \right)^{1/2} + 10^{-12}}.
\end{align}
For unseen rollout, we compare full predicted and true trajectory arrays on shared initial conditions and report the relative trajectory error
\begin{equation}
\varepsilon_{\mathrm{roll}}
=
\frac{\|u_{\mathrm{true}}-u_{\mathrm{pred}}\|_{2}}
{\|u_{\mathrm{true}}\|_{2}+10^{-12}}.
\end{equation}

The central question of the paper is not whether a model can reduce a training loss, but whether the recovered closure survives the full chain
\begin{equation}
\begin{aligned}
\text{observations}
\rightarrow {} & \text{numerical surrogate}
\rightarrow \text{symbolic compression} \\
\rightarrow {} & \text{forward validation}.
\end{aligned}
\end{equation}

\section{Method}

\subsection{Weak-form constitutive recovery}

Direct strong-form differentiation is notoriously brittle under noise and sparse observations. We therefore train the numerical closure surrogate in a \emph{space-weak} form. For a spatial test function $\phi=\phi(x)$ and periodic boundary conditions, \eqref{eq:rd_main} yields, for each observation time $t_n$,
\begin{equation}
\int_\Omega \phi u_t \, dx
+ \int_\Omega D(u) u_x \phi_x \, dx
- \int_\Omega \phi R(u) \, dx
= 0.
\label{eq:weak_form}
\end{equation}

Given a set of test functions $\{ \phi_k \}$ and discrete trajectories, we define the weak residual by projecting the observed field onto \eqref{eq:weak_form} at each saved time step. In the current implementation, the time derivative $u_t$ is approximated by a central difference on the saved temporal grid, while the spatial operator is handled in weak form. This avoids differentiating the data twice in space and provides a central physics-consistent component of the hybrid training signal. Related weak or integral formulations have been used to improve robustness of equation discovery under noisy or incomplete data \cite{Messenger_2021,Reinbold_2020,Tang_2023}.

\subsection{Numerical closure surrogate}

We use a compact neural surrogate with two scalar branches, one for diffusion and one for reaction:
\begin{equation}
\widehat{D}(u), \qquad \widehat{R}(u).
\end{equation}
The diffusion branch is constrained to remain positive by construction, while both branches are regularized for smoothness on the observed state range. In the current implementation the main backbone is a small residual MLP built on explicit polynomial features.

The training objective combines weak-form consistency with auxiliary rollout and regularization terms:
\begin{equation}
\mathcal{L}
= \mathcal{L}_{\text{weak}}
+ \alpha \mathcal{L}_{\text{roll}}
+ \beta \mathcal{L}_{\text{mass}}
+ \gamma \mathcal{L}_{\text{strong}}
+ \eta \mathcal{L}_{\text{anchor}}
+ \lambda \mathcal{L}_{\text{reg}}.
\label{eq11}
\end{equation}
Here $\mathcal{L}_{\text{mass}}$ explicitly reweights the spatially averaged balance relation already implied by the constant test function in the weak form, $\mathcal{L}_{\text{strong}}$ provides an auxiliary pointwise consistency term only in the synthetic regime, $\mathcal{L}_{\text{anchor}}$ softly anchors the reaction branch at the lower state boundary, and $\mathcal{L}_{\text{reg}}$ penalizes excessive variation of the learned closures. The conceptual pipeline remains weak-form driven: the auxiliary rollout, reweighted balance, anchor, and synthetic strong-form terms regularize the numerical surrogate, but the central identifiability question is still whether Stage~1 recovers the constitutive laws rather than merely fitting trajectories.

\subsection{Restricted symbolic compression}

Once the numerical closures are learned, we evaluate them on a dense support grid in $u$ and fit restricted symbolic families to the sampled constitutive curves. This stage is intentionally framed as \emph{symbolic compression of learned constitutive surrogates}, not as unrestricted symbolic discovery directly from raw trajectories.

The candidate families are deliberately small and interpretable. For diffusion we consider low-order polynomials, a low-order rational family, exponential-decay forms, and saturation forms. For reaction we consider low-order polynomials, a low-order rational family, and two one-factor nonlinear families of the forms $u\exp(\cdot)$ and $u/(1+\cdot)$. For each closure, we select the lowest-complexity candidate whose fit error lies within a small tolerance of the best candidate. This gives a symbolic pair
\begin{equation}
f_{\text{sym}} = \left( D_{\text{sym}}, R_{\text{sym}} \right)
\end{equation}
that is compact enough to inspect while remaining faithful to the neural surrogate.

This restricted library is a feature, not a bug. The goal of Stage~2 is not exhaustive symbolic search over arbitrary formulas. The goal is to test whether a learned numerical constitutive surrogate can be compressed into a compact family without materially changing its induced dynamics.

\subsection{Forward validation}

The symbolic stage only counts if the compressed closure can be reinserted into the PDE and rolled forward successfully. We therefore validate both the learned numerical surrogate and the compressed symbolic closure by forward simulation on shared unseen initial conditions. This produces two rollout metrics:
\begin{equation}
\varepsilon_{\text{roll}}^{\text{neural}},
\qquad
\varepsilon_{\text{roll}}^{\text{sym}}.
\end{equation}
Their comparison is essential: if the symbolic closure substantially degrades rollout, then the compression stage failed even if curve fitting looked accurate in state space.

\section{Theory}

The theory in this paper is intended as a mechanism-level explanation of the observed behavior, not as a full uniqueness-and-convergence theory for nonlinear PDE inverse problems. Its role is to justify the weak-form objective, explain when polynomial baselines are especially strong in the correctly specified regime, and formalize why symbolic compression preserves but does not repair constitutive bias.

Unless stated otherwise, the closure-space norms below are taken in a common norm on the scalar constitutive functions. When a norm is restricted to the bounded state interval visited by the trajectories, we write it explicitly as an $L^\infty(\mathcal{U})$ norm. In the empirical evaluation, the reported closure metrics are relative discrete $L^2(\mathcal{U})$ errors on a common support grid, which is consistent with the topology induced by the weak-form objective and the rollout comparisons used in the paper.

\subsection{Weak-form consistency}

\begin{proposition}[Space-weak consistency]
Assume
\begin{equation}
u \in L^2(0,T;H^1(\Omega)),
\qquad
u_t \in L^2(0,T;H^{-1}(\Omega)).
\end{equation}
Then for any spatial test function $\phi \in H^1(\Omega)$,
\begin{equation}
\int_0^T \langle u_t,\phi \rangle \, dt
+ \int_0^T \int_\Omega D(u) u_x \phi_x \, dx dt
- \int_0^T \int_\Omega R(u) \phi \, dx dt
= 0.
\end{equation}
Under periodic or no-flux boundaries,
\begin{equation}
\frac{d}{dt} \int_\Omega u \, dx = \int_\Omega R(u) \, dx.
\end{equation}
\end{proposition}

This proposition legitimizes the weak residual and the reweighted mass-balance penalty. The second identity is obtained by choosing $\phi \equiv 1$, so the diffusion contribution vanishes and only the reaction-driven mass balance remains. In the implementation, $\mathcal{L}_{\text{mass}}$ isolates this scalar relation and gives it additional weight on top of the general weak residual. The result should be read as a consistency statement for the space-weak training objective. It is not intended as a complete space-time weak-solution definition.

\subsection{Matched-library identification}

\begin{proposition}[Discrete projected identification under sufficient excitation]
Suppose the true closures lie in finite dictionaries:
\begin{equation}
D(u) = \sum_{i=1}^{p_D} a_i \psi_i(u),
\qquad
R(u) = \sum_{j=1}^{p_R} b_j \chi_j(u).
\end{equation}
After weak-form projection and discrete quadrature, the identification problem becomes
\begin{equation}
\Phi \theta = Y,
\end{equation}
where $\theta$ stacks the coefficients. If the Gram matrix
\begin{equation}
G = \Phi^\top \Phi
\end{equation}
is positive definite, then the least-squares estimator is unique.
\end{proposition}

Proof sketch. The discrete projected fit is the least-squares problem
\begin{equation}
\min_\theta \|\Phi\theta - Y\|_2^2.
\end{equation}
Its normal equations are $G\theta = \Phi^\top Y$ with $G=\Phi^\top\Phi$. If $G$ is positive definite, then it is invertible, so the least-squares estimator is unique and equals $\theta = G^{-1}\Phi^\top Y$.

This explains why weak polynomial baselines are so strong in Case~A and Case~B. The important point is not merely that the model is linear in coefficients, but that excitation coverage determines whether the induced design matrix has informative, nondegenerate columns. This is a uniqueness statement for the projected coefficient fit, not a full statistical consistency theorem for PDE closure recovery.

\subsection{Low residual does not imply true closure recovery}

\begin{proposition}[Non-identifiability under limited excitation]
\label{prop:non_identifiability}
Let $\mathcal{A}_{\mathcal{T}}(D,R)$ denote the weak-form observation operator induced by a finite trajectory set $\mathcal{T}$. If $\mathcal{A}_{\mathcal{T}}$ is not injective on the admissible closure class, then there exist nonzero perturbations $(\delta D,\delta R)$ such that
\begin{equation}
\mathcal{A}_{\mathcal{T}}(D+\delta D,R+\delta R)
\approx
\mathcal{A}_{\mathcal{T}}(D,R).
\end{equation}
Hence low weak residual on the observed trajectories does not imply unique or correct recovery of the true closure pair.
\end{proposition}

Here ``$\approx$'' means approximate equality in the induced observation norm or, equivalently in practice, a comparably small weak residual on the observed trajectory set. This proposition formalizes the central inverse-problem warning behind the paper. It explains why low weak loss and small short-horizon rollout error can coexist with large closure error, especially when the observed data occupy only a narrow state interval or have weak gradient and curvature content.

\subsection{Approximation bias under function-class mismatch}

\begin{proposition}[Approximation bias]
Let $\mathcal{F}$ be the hypothesis class used by the identifier. If the true closure $f^\ast$ does not belong to $\mathcal{F}$, define the best-in-class target
\begin{equation}
f^\dagger = \arg \min_{f \in \mathcal{F}} \mathcal{L}(f).
\end{equation}
Then the recovery error satisfies
\begin{equation}
\|f^\ast - \widehat{f}\|
\le
\|f^\ast - f^\dagger\|
+ \|f^\dagger - \widehat{f}\|.
\end{equation}
\end{proposition}

The first term is model-class bias and the second is identification or optimization error. This decomposition explains why the mismatch regime is the right place to justify a neural surrogate: its main value is reduced approximation bias relative to a misspecified restricted library.

\subsection{Symbolic compression as bias-preserving distillation}

\begin{proposition}[Bias preservation under symbolic compression]
Let $f^\ast$ be the true closure, $\widehat{f}$ the learned numerical surrogate, and $f_{\text{sym}}$ the compressed symbolic closure. Then
\begin{equation}
\left| \|f^\ast - f_{\text{sym}}\| - \|f^\ast - \widehat{f}\| \right|
\le
\|\widehat{f} - f_{\text{sym}}\|.
\label{eq:bias_preservation}
\end{equation}
\end{proposition}

Proof sketch. Apply the reverse triangle inequality to $a=f^\ast-f_{\text{sym}}$ and $b=f^\ast-\widehat{f}$:
\begin{equation}
\bigl| \|a\| - \|b\| \bigr| \le \|a-b\| = \|\widehat{f}-f_{\text{sym}}\|,
\end{equation}
which is exactly \eqref{eq:bias_preservation}.

Equation~\eqref{eq:bias_preservation} is the core theoretical explanation of the benchmark results. If compression error is small, then the symbolic true error must remain close to the neural true error. In other words, the symbolic stage cannot create a large correction unless it first departs substantially from the surrogate it is meant to compress.
The estimate is a direct application of the reverse triangle inequality to the pair of vectors $f^\ast-f_{\text{sym}}$ and $f^\ast-\widehat{f}$.

\subsection{Rollout preservation under small closure perturbations}

\begin{remark}[Finite-time rollout stability]
Let $u_1$ and $u_2$ solve the same initial-value problem on $[0,T]$ with closure pairs $f_1=(D_1,R_1)$ and $f_2=(D_2,R_2)$. Under uniform parabolicity and bounded-Lipschitz assumptions on both closures, there exists a constant $C_T$ such that
\begin{equation}
\sup_{t \in [0,T]} \|u_1(t) - u_2(t)\|_{L^2}
\le
C_T \left(
\|D_1-D_2\|_{L^\infty(\mathcal{U})} + \|R_1-R_2\|_{L^\infty(\mathcal{U})}
\right),
\end{equation}
where $\mathcal{U}$ is the bounded state range visited by the trajectories.
\end{remark}

This estimate should be read as a standard finite-time stability template rather than a fully derived theorem under minimal assumptions. It connects small closure-space compression error to small additional rollout degradation over finite horizons and explains why symbolic rollout almost matches neural rollout in the experiments.

\subsection{Unified error decomposition}

Combining the approximation-bias and compression statements gives the practical decomposition
\begin{equation}
\begin{aligned}
\|f^\ast - f_{\text{sym}}\|
\le {} &
\underbrace{\|f^\ast - f^\dagger\|}_{\text{model-class bias}}
\\
&+
\underbrace{\|f^\dagger - \widehat{f}\|}_{\text{identification and optimization error}}
\\
&+
\underbrace{\|\widehat{f} - f_{\text{sym}}\|}_{\text{symbolic compression error}}.
\end{aligned}
\label{eq:total_decomposition}
\end{equation}
The experiments in this paper indicate that, on the reported benchmarks, the third term is comparatively small while the first two dominate. On these benchmarks, this is why the symbolic stage is not the main bottleneck.

\section{Experiments}

\subsection{Experimental protocol}

All main experiments are performed on 1D periodic reaction-diffusion systems. We evaluate three representative cases:
\begin{itemize}[leftmargin=1.5em]
\item \textbf{Case A:} a matched-library reference case with
\begin{equation*}
D(u) = 0.01 + 0.05u, \qquad R(u) = u(1-u).
\end{equation*}
\item \textbf{Case B:} a second matched-library case with
\begin{equation*}
D(u) = 0.01 + 0.03u^2, \qquad R(u) = u - 1.5u^2 + 0.5u^3.
\end{equation*}
\item \textbf{Case Exp:} a function-class-mismatch stress test with
\begin{equation*}
D(u) = 0.01 + 0.035(1-e^{-2.5u}), \qquad R(u) = u(1.1e^{-1.4u} - 0.22).
\end{equation*}
\end{itemize}

We compare four method families:
\begin{enumerate}[leftmargin=1.5em]
\item strong polynomial baseline,
\item weak polynomial baseline,
\item neural numerical surrogate,
\item neural surrogate followed by symbolic compression.
\end{enumerate}
Unless otherwise stated, the neural benchmarks use $8$ training trajectories, $4$-mode random Fourier initial conditions, and the simulation grid $n_x=64$ with integrator step size $\Delta t=10^{-4}$ up to final time $T=0.1$, with snapshots saved every $10$ steps. Models are trained for $150$ epochs. A ``5\% noise'' setting means additive Gaussian perturbations with standard deviation $0.05\,\mathrm{std}(u)$. A ``sparse $x+t$'' setting means spatial stride $2$ and temporal stride $2$. Each reported value is averaged across three random seeds.

\subsection{Baseline map: matched library versus mismatch}

Table~\ref{tab:baseline_map} summarizes representative settings. The main qualitative pattern is immediate. In the matched-library regime, strong and weak polynomial baselines are extremely competitive and behave like correctly specified reference estimators. In the mismatch regime, the neural surrogate becomes relevant as a flexible numerical approximation layer, but its advantage is metric-dependent and does not automatically translate into better rollout. In fact, for Case~Exp under clean data, the weak polynomial baseline still gives the best rollout and the smallest reaction error, while the strong polynomial baseline remains competitive on diffusion. Function-class mismatch therefore does not by itself guarantee a uniform neural advantage under the current Stage~1 objective.

\begin{table}[t]
\centering
\small
\resizebox{\textwidth}{!}{%
\begin{tabular}{lllccc}
\toprule
Case & Setting & Method & ErrD & ErrR & Unseen \\
\midrule
Case A & clean & strong\_poly & $4.098\mathrm{e}{-04} \pm 1.144\mathrm{e}{-04}$ & $2.113\mathrm{e}{-03} \pm 8.855\mathrm{e}{-04}$ & $1.804\mathrm{e}{-05} \pm 5.128\mathrm{e}{-06}$ \\
Case A & clean & weak\_poly & $8.151\mathrm{e}{-03} \pm 1.047\mathrm{e}{-03}$ & $3.090\mathrm{e}{-02} \pm 9.034\mathrm{e}{-03}$ & $8.307\mathrm{e}{-04} \pm 1.348\mathrm{e}{-04}$ \\
Case A & clean & neural & $6.926\mathrm{e}{-02} \pm 4.007\mathrm{e}{-03}$ & $3.128\mathrm{e}{-01} \pm 1.892\mathrm{e}{-02}$ & $4.828\mathrm{e}{-03} \pm 3.137\mathrm{e}{-04}$ \\
Case A & clean & neural+symbolic & $6.853\mathrm{e}{-02} \pm 4.266\mathrm{e}{-03}$ & $3.127\mathrm{e}{-01} \pm 1.894\mathrm{e}{-02}$ & $4.701\mathrm{e}{-03} \pm 3.516\mathrm{e}{-04}$ \\
\midrule
Case B & clean & strong\_poly & $4.018\mathrm{e}{-04} \pm 1.088\mathrm{e}{-04}$ & $2.597\mathrm{e}{-03} \pm 1.092\mathrm{e}{-03}$ & $1.651\mathrm{e}{-05} \pm 4.630\mathrm{e}{-06}$ \\
Case B & clean & weak\_poly & $8.030\mathrm{e}{-03} \pm 6.990\mathrm{e}{-04}$ & $3.701\mathrm{e}{-02} \pm 8.150\mathrm{e}{-03}$ & $9.556\mathrm{e}{-04} \pm 1.252\mathrm{e}{-04}$ \\
Case B & clean & neural & $1.003\mathrm{e}{-01} \pm 5.687\mathrm{e}{-03}$ & $2.047\mathrm{e}{-01} \pm 1.727\mathrm{e}{-02}$ & $3.949\mathrm{e}{-03} \pm 3.374\mathrm{e}{-04}$ \\
Case B & clean & neural+symbolic & $1.001\mathrm{e}{-01} \pm 5.650\mathrm{e}{-03}$ & $2.036\mathrm{e}{-01} \pm 1.756\mathrm{e}{-02}$ & $3.727\mathrm{e}{-03} \pm 3.585\mathrm{e}{-04}$ \\
\midrule
Case Exp & clean & strong\_poly & $4.609\mathrm{e}{-02} \pm 2.701\mathrm{e}{-04}$ & $4.270\mathrm{e}{-01} \pm 1.115\mathrm{e}{-01}$ & $3.146\mathrm{e}{-03} \pm 2.813\mathrm{e}{-04}$ \\
Case Exp & clean & weak\_poly & $5.230\mathrm{e}{-02} \pm 1.786\mathrm{e}{-03}$ & $8.712\mathrm{e}{-02} \pm 1.598\mathrm{e}{-02}$ & $2.738\mathrm{e}{-03} \pm 8.345\mathrm{e}{-05}$ \\
Case Exp & clean & neural & $4.772\mathrm{e}{-02} \pm 7.372\mathrm{e}{-04}$ & $4.400\mathrm{e}{-01} \pm 8.150\mathrm{e}{-03}$ & $3.212\mathrm{e}{-03} \pm 7.913\mathrm{e}{-05}$ \\
Case Exp & clean & neural+symbolic & $4.765\mathrm{e}{-02} \pm 8.388\mathrm{e}{-04}$ & $4.400\mathrm{e}{-01} \pm 8.151\mathrm{e}{-03}$ & $3.226\mathrm{e}{-03} \pm 9.908\mathrm{e}{-05}$ \\
\bottomrule
\end{tabular}
}
\caption{Representative baseline comparison. In matched-library settings, polynomial baselines are exceptionally strong. In mismatch settings, the neural surrogate serves as a flexible numerical constitutive approximator, while the symbolic stage largely preserves its behavior.}
\label{tab:baseline_map}
\end{table}

\subsection{Excitation coverage and identifiability}

Table~\ref{tab:excitation} makes the identifiability issue explicit. For computational efficiency, this excitation benchmark uses a lighter protocol with $n_x=48$, $T=0.05$, snapshots saved every $5$ steps, and $6$ training trajectories. We compare low-excitation and high-excitation training sets by shrinking or expanding the amplitude range of the random initial conditions. The low-excitation regime occupies only a narrow state interval and produces much weaker gradient and diffusion-energy statistics. In all cases, this regime attains a \emph{smaller} weak loss and even a substantially smaller short-horizon unseen rollout error, yet the recovered closures are substantially worse. Because the signal energy is also much smaller in the low-excitation regime, these weak-loss values should be read as absolute residual magnitudes rather than normalized cross-dataset scores. The conclusion is the one predicted by Proposition~\ref{prop:non_identifiability}: low residual is not enough when the observation operator is weakly excited.

\begin{table}[t]
\centering
\small
\resizebox{\textwidth}{!}{%
\begin{tabular}{llcccccc}
\toprule
Case & Regime & Bin coverage & Weak diffusion energy & Weak loss & ErrD & ErrR & Unseen \\
\midrule
Case A & low excitation & $1.250\mathrm{e}{-01} \pm 0.000\mathrm{e}{+00}$ & $1.091\mathrm{e}{+00} \pm 1.270\mathrm{e}{-01}$ & $1.879\mathrm{e}{-05} \pm 3.018\mathrm{e}{-06}$ & $1.280\mathrm{e}{-01} \pm 9.278\mathrm{e}{-03}$ & $1.289\mathrm{e}{+00} \pm 3.130\mathrm{e}{-02}$ & $5.333\mathrm{e}{-04} \pm 8.114\mathrm{e}{-05}$ \\
Case A & high excitation & $1.000\mathrm{e}{+00} \pm 0.000\mathrm{e}{+00}$ & $1.158\mathrm{e}{+02} \pm 1.368\mathrm{e}{+01}$ & $9.471\mathrm{e}{-04} \pm 1.602\mathrm{e}{-04}$ & $5.488\mathrm{e}{-02} \pm 1.125\mathrm{e}{-02}$ & $3.673\mathrm{e}{-01} \pm 2.049\mathrm{e}{-02}$ & $3.879\mathrm{e}{-03} \pm 3.066\mathrm{e}{-04}$ \\
\midrule
Case B & low excitation & $1.250\mathrm{e}{-01} \pm 0.000\mathrm{e}{+00}$ & $1.822\mathrm{e}{+00} \pm 2.590\mathrm{e}{-01}$ & $1.616\mathrm{e}{-05} \pm 2.869\mathrm{e}{-06}$ & $2.984\mathrm{e}{-01} \pm 1.291\mathrm{e}{-02}$ & $1.031\mathrm{e}{+00} \pm 8.982\mathrm{e}{-02}$ & $4.744\mathrm{e}{-04} \pm 8.218\mathrm{e}{-05}$ \\
Case B & high excitation & $1.000\mathrm{e}{+00} \pm 0.000\mathrm{e}{+00}$ & $2.046\mathrm{e}{+02} \pm 2.852\mathrm{e}{+01}$ & $3.691\mathrm{e}{-04} \pm 5.804\mathrm{e}{-05}$ & $8.566\mathrm{e}{-02} \pm 4.676\mathrm{e}{-03}$ & $2.829\mathrm{e}{-01} \pm 3.448\mathrm{e}{-02}$ & $3.310\mathrm{e}{-03} \pm 1.639\mathrm{e}{-04}$ \\
\midrule
Case Exp & low excitation & $1.250\mathrm{e}{-01} \pm 0.000\mathrm{e}{+00}$ & $2.411\mathrm{e}{+00} \pm 2.782\mathrm{e}{-01}$ & $1.259\mathrm{e}{-05} \pm 1.876\mathrm{e}{-06}$ & $1.706\mathrm{e}{-01} \pm 1.758\mathrm{e}{-02}$ & $8.154\mathrm{e}{-01} \pm 1.803\mathrm{e}{-02}$ & $2.743\mathrm{e}{-04} \pm 1.822\mathrm{e}{-05}$ \\
Case Exp & high excitation & $1.000\mathrm{e}{+00} \pm 0.000\mathrm{e}{+00}$ & $2.376\mathrm{e}{+02} \pm 2.818\mathrm{e}{+01}$ & $3.872\mathrm{e}{-04} \pm 7.170\mathrm{e}{-05}$ & $4.657\mathrm{e}{-02} \pm 1.591\mathrm{e}{-03}$ & $4.052\mathrm{e}{-01} \pm 2.217\mathrm{e}{-02}$ & $2.671\mathrm{e}{-03} \pm 4.586\mathrm{e}{-05}$ \\
\bottomrule
\end{tabular}
}
\caption{Excitation benchmark with three seeds. Narrow low-excitation training sets achieve very small weak losses but much worse closure recovery. The observed trajectory fit alone is therefore not a reliable identifiability diagnostic.}
\label{tab:excitation}
\end{table}

\subsection{Symbolic compression preserves surrogate behavior}

\begin{figure}[t]
\centering
\includegraphics[width=\textwidth]{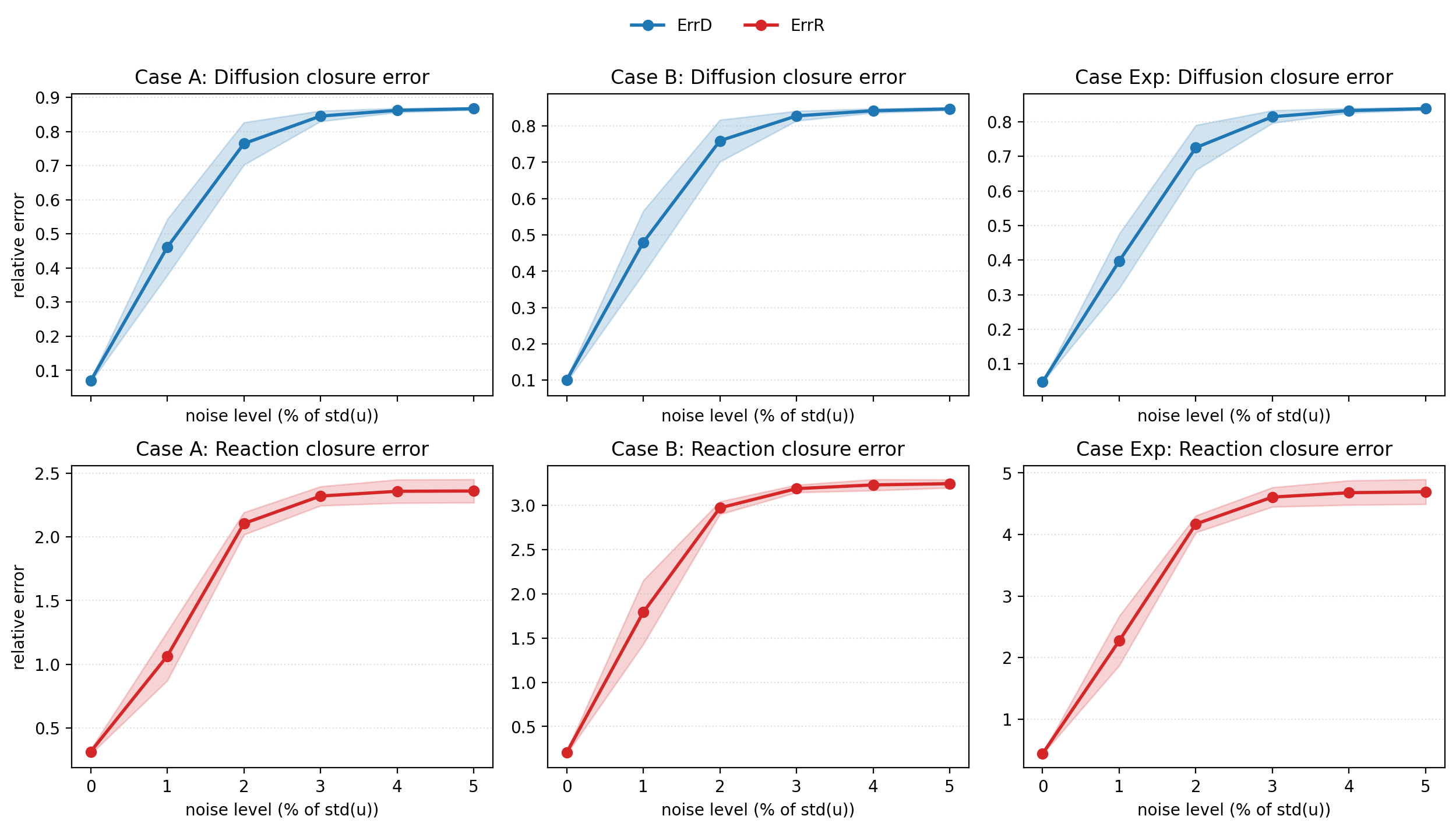}
\caption{Stage~1 noise-sensitivity sweep across the three 1D benchmark cases. Mean$\pm$standard deviation over three seeds for the neural Stage~1 closure errors. In all cases, both diffusion and reaction recovery deteriorate sharply between $1\%$ and $3\%$ observation noise and then flatten toward saturated high-error regimes by $5\%$.}
\label{fig:noise_sweep_all_cases}
\end{figure}

\begin{figure}[t]
\centering
\includegraphics[width=\textwidth]{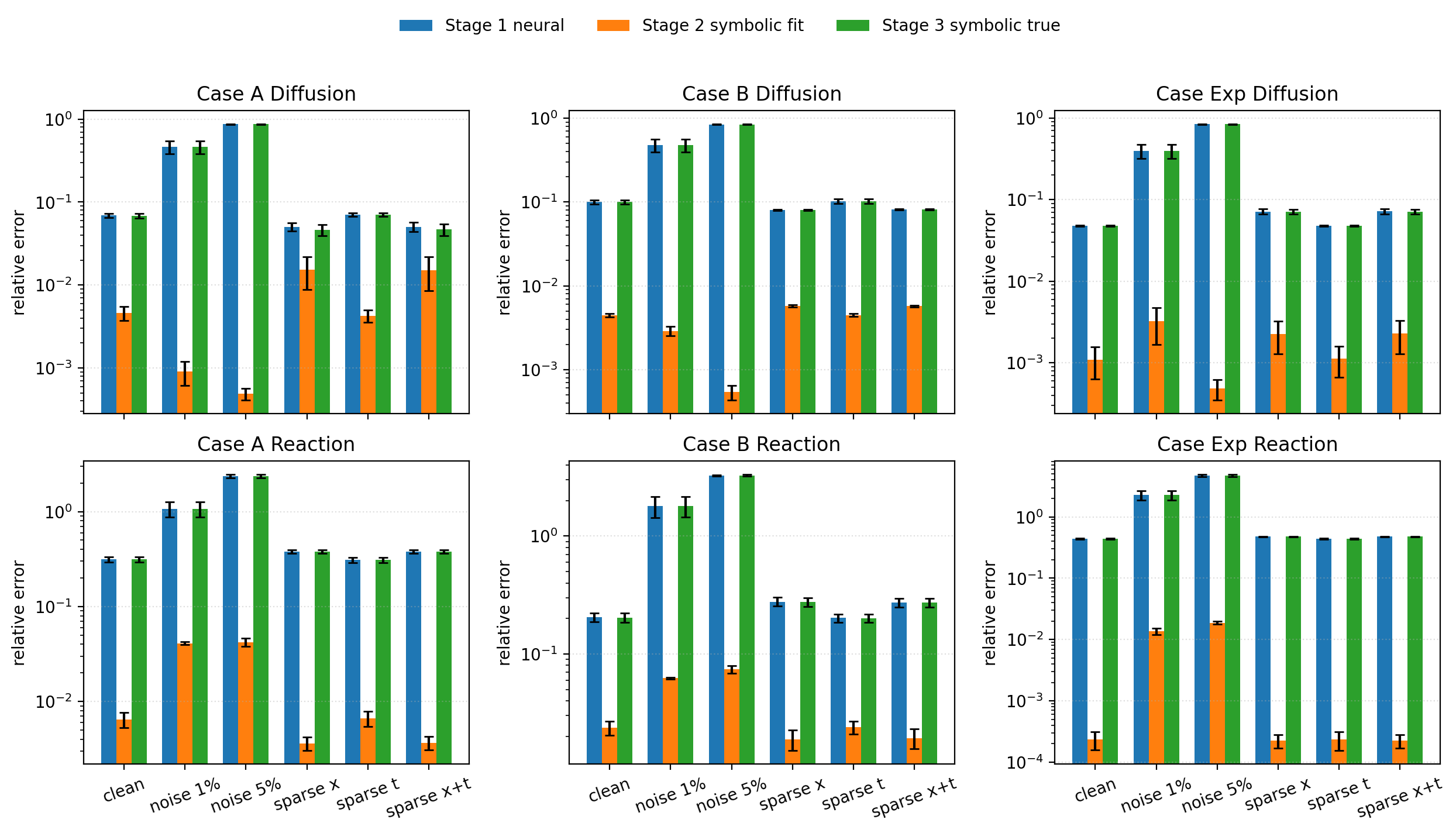}
\caption{Neural-to-symbolic error propagation. Stage~2 compression error remains small, while Stage~3 true symbolic error tracks the Stage~1 neural true error. This is the empirical signature of bias preservation.}
\label{fig:error_propagation}
\end{figure}

The core symbolic benchmark is summarized in Table~\ref{tab:symbolic_summary} and Figure~\ref{fig:error_propagation}. Across the reported settings, the symbolic surrogate error remains much smaller than the neural true error. This indicates that the symbolic stage is primarily compressing the learned constitutive curves rather than inventing a different law. At the same time, the ``noise 5\%'' rows indicate that the current weak-form-driven hybrid Stage~1 objective is not robust to this noise level on the present benchmarks: the neural true closure errors deteriorate sharply under observation noise, while the symbolic stage continues to inherit that constitutive bias rather than repair it.
A dedicated six-level noise sweep across all three 1D cases (Figure~\ref{fig:noise_sweep_all_cases}) shows that these $5\%$ rows are not isolated outliers. In each of the three reported cases, the mean closure errors rise sharply already between $1\%$ and $3\%$ noise and then flatten toward the $5\%$ regime, so the practically relevant phenomenon is a protocol-specific degradation knee rather than a single pathological endpoint.

To quantify bias propagation, we define the bias inheritance ratios
\begin{equation}
\mathrm{BIR}_D = \frac{\mathrm{ErrD}^{\mathrm{sym,true}}}{\mathrm{ErrD}^{\mathrm{neural,true}}},
\qquad
\mathrm{BIR}_R = \frac{\mathrm{ErrR}^{\mathrm{sym,true}}}{\mathrm{ErrR}^{\mathrm{neural,true}}}.
\end{equation}
Values near one mean that symbolic compression preserves the constitutive bias already present in the neural surrogate. Values modestly below one indicate mild smoothing, not a systematic correction mechanism.

\begin{table}[t]
\centering
\small
\resizebox{\textwidth}{!}{%
\begin{tabular}{llcccccc}
\toprule
Case & Setting & Neural ErrD & Neural ErrR & Sym surrogate ErrD & Sym surrogate ErrR & $\mathrm{BIR}_D$ & $\mathrm{BIR}_R$ \\
\midrule
Case A & clean & $6.926\mathrm{e}{-02}$ & $3.128\mathrm{e}{-01}$ & $4.628\mathrm{e}{-03}$ & $6.453\mathrm{e}{-03}$ & $9.893\mathrm{e}{-01}$ & $9.999\mathrm{e}{-01}$ \\
Case A & noise 5\% & $8.672\mathrm{e}{-01}$ & $2.361\mathrm{e}{+00}$ & $4.910\mathrm{e}{-04}$ & $4.212\mathrm{e}{-02}$ & $1.000\mathrm{e}{+00}$ & $1.002\mathrm{e}{+00}$ \\
Case A & sparse x+t & $5.046\mathrm{e}{-02}$ & $3.786\mathrm{e}{-01}$ & $1.518\mathrm{e}{-02}$ & $3.676\mathrm{e}{-03}$ & $9.239\mathrm{e}{-01}$ & $1.000\mathrm{e}{+00}$ \\
\midrule
Case B & clean & $1.003\mathrm{e}{-01}$ & $2.047\mathrm{e}{-01}$ & $4.431\mathrm{e}{-03}$ & $2.370\mathrm{e}{-02}$ & $9.988\mathrm{e}{-01}$ & $9.942\mathrm{e}{-01}$ \\
Case B & noise 5\% & $8.470\mathrm{e}{-01}$ & $3.248\mathrm{e}{+00}$ & $5.387\mathrm{e}{-04}$ & $7.419\mathrm{e}{-02}$ & $1.000\mathrm{e}{+00}$ & $1.007\mathrm{e}{+00}$ \\
Case B & sparse x+t & $8.184\mathrm{e}{-02}$ & $2.728\mathrm{e}{-01}$ & $5.720\mathrm{e}{-03}$ & $1.937\mathrm{e}{-02}$ & $9.969\mathrm{e}{-01}$ & $9.981\mathrm{e}{-01}$ \\
\midrule
Case Exp & clean & $4.772\mathrm{e}{-02}$ & $4.400\mathrm{e}{-01}$ & $1.099\mathrm{e}{-03}$ & $2.363\mathrm{e}{-04}$ & $9.985\mathrm{e}{-01}$ & $1.000\mathrm{e}{+00}$ \\
Case Exp & noise 5\% & $8.379\mathrm{e}{-01}$ & $4.695\mathrm{e}{+00}$ & $4.861\mathrm{e}{-04}$ & $1.877\mathrm{e}{-02}$ & $1.000\mathrm{e}{+00}$ & $9.992\mathrm{e}{-01}$ \\
Case Exp & sparse x+t & $7.167\mathrm{e}{-02}$ & $4.759\mathrm{e}{-01}$ & $2.282\mathrm{e}{-03}$ & $2.231\mathrm{e}{-04}$ & $9.987\mathrm{e}{-01}$ & $1.000\mathrm{e}{+00}$ \\
\bottomrule
\end{tabular}
}
\caption{Representative symbolic compression summary. The symbolic stage stays close to the neural surrogate and the bias inheritance ratios remain near one, especially on the mismatch stress test.}
\label{tab:symbolic_summary}
\end{table}

\subsection{PySR replacement check}

To test whether bias inheritance is merely an artifact of the restricted Stage~2 family, we replaced the symbolic-compression step with PySR \cite{Cranmer_2023_PySR} on the mismatch stress test (Case~Exp) under the clean and ``noise 5\%'' settings. The goal of this check is not to claim that PySR is the canonical Stage~2 engine, but to ask a sharper question: if Stage~2 is made substantially more expressive, does it actually repair the constitutive bias inherited from Stage~1? Table~\ref{tab:pysr_check} indicates that, on this check, the answer remains no.

In the clean setting, PySR attains a small surrogate-fit error and a rollout of the same order as the restricted symbolic family, yet its true closure errors remain essentially inherited from the neural surrogate, with $\mathrm{BIR}_D \approx 9.985\times 10^{-1}$ and $\mathrm{BIR}_R \approx 9.958\times 10^{-1}$. Under $5\%$ noise, the conclusion becomes even sharper on this mismatch stress test: both the restricted family and PySR remain at $\mathrm{BIR}\approx 1$, and their rollout errors stay near $1.04\times 10^{-1}$. The main differences are complexity and, in the clean setting, somewhat looser surrogate compression rather than systematic correction. Averaged over three seeds, the PySR expressions have mean complexities of roughly $(17, 20.7)$ in the clean setting and $(17, 22)$ under $5\%$ noise, versus $(5,5)$ and $(3,5)$ for the restricted family. This is consistent with the mechanism claim of the paper on the tested mismatch settings: bias inheritance is not caused solely by a weak symbolic library; it persists even when Stage~2 is replaced by a stronger symbolic search engine.

\begin{table}[t]
\centering
\small
\resizebox{\textwidth}{!}{%
\begin{tabular}{llccccc}
\toprule
Setting & Stage 2 & Surrogate ErrD & Surrogate ErrR & Unseen & $\mathrm{BIR}_D$ & $\mathrm{BIR}_R$ \\
\midrule
clean & restricted & $1.099\mathrm{e}{-03} \pm 4.705\mathrm{e}{-04}$ & $2.363\mathrm{e}{-04} \pm 7.787\mathrm{e}{-05}$ & $3.226\mathrm{e}{-03} \pm 9.908\mathrm{e}{-05}$ & $9.985\mathrm{e}{-01} \pm 2.316\mathrm{e}{-03}$ & $1.000\mathrm{e}{+00} \pm 1.649\mathrm{e}{-06}$ \\
clean & PySR & $5.638\mathrm{e}{-03} \pm 2.965\mathrm{e}{-03}$ & $7.764\mathrm{e}{-03} \pm 5.431\mathrm{e}{-03}$ & $3.302\mathrm{e}{-03} \pm 1.041\mathrm{e}{-04}$ & $9.985\mathrm{e}{-01} \pm 6.270\mathrm{e}{-03}$ & $9.958\mathrm{e}{-01} \pm 4.358\mathrm{e}{-03}$ \\
\midrule
noise 5\% & restricted & $4.861\mathrm{e}{-04} \pm 1.349\mathrm{e}{-04}$ & $1.877\mathrm{e}{-02} \pm 9.893\mathrm{e}{-04}$ & $1.038\mathrm{e}{-01} \pm 2.461\mathrm{e}{-03}$ & $1.000\mathrm{e}{+00} \pm 3.239\mathrm{e}{-06}$ & $9.992\mathrm{e}{-01} \pm 3.095\mathrm{e}{-04}$ \\
noise 5\% & PySR & $5.019\mathrm{e}{-04} \pm 1.329\mathrm{e}{-04}$ & $2.161\mathrm{e}{-02} \pm 3.547\mathrm{e}{-03}$ & $1.038\mathrm{e}{-01} \pm 2.499\mathrm{e}{-03}$ & $1.000\mathrm{e}{+00} \pm 3.726\mathrm{e}{-06}$ & $1.000\mathrm{e}{+00} \pm 5.670\mathrm{e}{-04}$ \\
\bottomrule
\end{tabular}
}
\caption{Case~Exp Stage~2 replacement check. On this mismatch stress test, replacing the restricted symbolic family with PySR leaves the qualitative conclusion unchanged: Stage~2 still preserves the constitutive bias inherited from the neural surrogate, especially under observation noise.}
\label{tab:pysr_check}
\end{table}

\subsection{Cross-resolution anti-inverse-crime check}

Table~\ref{tab:cross_resolution} evaluates the learned neural closure on a validation solver with a different grid and saved time step from the data-generation solver. The same-grid unseen rollout and the cross-grid rollout remain very close in all four observation settings. This does not prove full discretization independence, but it does indicate that the reported mismatch conclusions are not a trivial artifact of reusing exactly the same numerical configuration at training and validation time. A second pattern is also visible: in this benchmark, spatial downsampling is substantially more damaging than temporal downsampling, while temporal stride $2$ alone leaves the errors almost unchanged relative to the $(1,1)$ reference.

\begin{table}[t]
\centering
\small
\resizebox{\textwidth}{!}{%
\begin{tabular}{lccccc}
\toprule
Observed stride $(x,t)$ & ErrD & ErrR & Weak loss & Unseen & Cross-grid \\
\midrule
$(1,1)$ & $4.585\mathrm{e}{-02} \pm 5.543\mathrm{e}{-04}$ & $4.081\mathrm{e}{-01} \pm 1.701\mathrm{e}{-02}$ & $3.172\mathrm{e}{-04} \pm 1.145\mathrm{e}{-05}$ & $2.606\mathrm{e}{-03} \pm 2.861\mathrm{e}{-05}$ & $2.597\mathrm{e}{-03} \pm 3.773\mathrm{e}{-05}$ \\
$(2,1)$ & $6.838\mathrm{e}{-02} \pm 5.635\mathrm{e}{-03}$ & $4.377\mathrm{e}{-01} \pm 2.040\mathrm{e}{-02}$ & $1.126\mathrm{e}{-03} \pm 4.381\mathrm{e}{-04}$ & $3.678\mathrm{e}{-03} \pm 2.677\mathrm{e}{-04}$ & $3.675\mathrm{e}{-03} \pm 2.602\mathrm{e}{-04}$ \\
$(1,2)$ & $4.584\mathrm{e}{-02} \pm 5.746\mathrm{e}{-04}$ & $4.082\mathrm{e}{-01} \pm 1.685\mathrm{e}{-02}$ & $3.173\mathrm{e}{-04} \pm 1.087\mathrm{e}{-05}$ & $2.606\mathrm{e}{-03} \pm 2.908\mathrm{e}{-05}$ & $2.597\mathrm{e}{-03} \pm 3.842\mathrm{e}{-05}$ \\
$(2,2)$ & $6.839\mathrm{e}{-02} \pm 5.664\mathrm{e}{-03}$ & $4.378\mathrm{e}{-01} \pm 2.029\mathrm{e}{-02}$ & $1.126\mathrm{e}{-03} \pm 4.383\mathrm{e}{-04}$ & $3.678\mathrm{e}{-03} \pm 2.689\mathrm{e}{-04}$ & $3.676\mathrm{e}{-03} \pm 2.614\mathrm{e}{-04}$ \\
\bottomrule
\end{tabular}
}
\caption{Cross-resolution anti-inverse-crime benchmark on Case~Exp. This benchmark uses a separately generated dataset with a different grid and snapshot cadence from the default protocol. Training data are generated on a fine solver ($n_x=64$, saved $\Delta t=5\times 10^{-4}$, saved horizon $5.0\times 10^{-2}$), while validation uses a different grid and saved time step ($n_x=48$, saved $\Delta t=6\times 10^{-4}$, last saved snapshot at $4.98\times 10^{-2}$).}
\label{tab:cross_resolution}
\end{table}

\subsection{Rollout preservation and bias inheritance}

\begin{figure}[t]
\centering
\includegraphics[width=\textwidth]{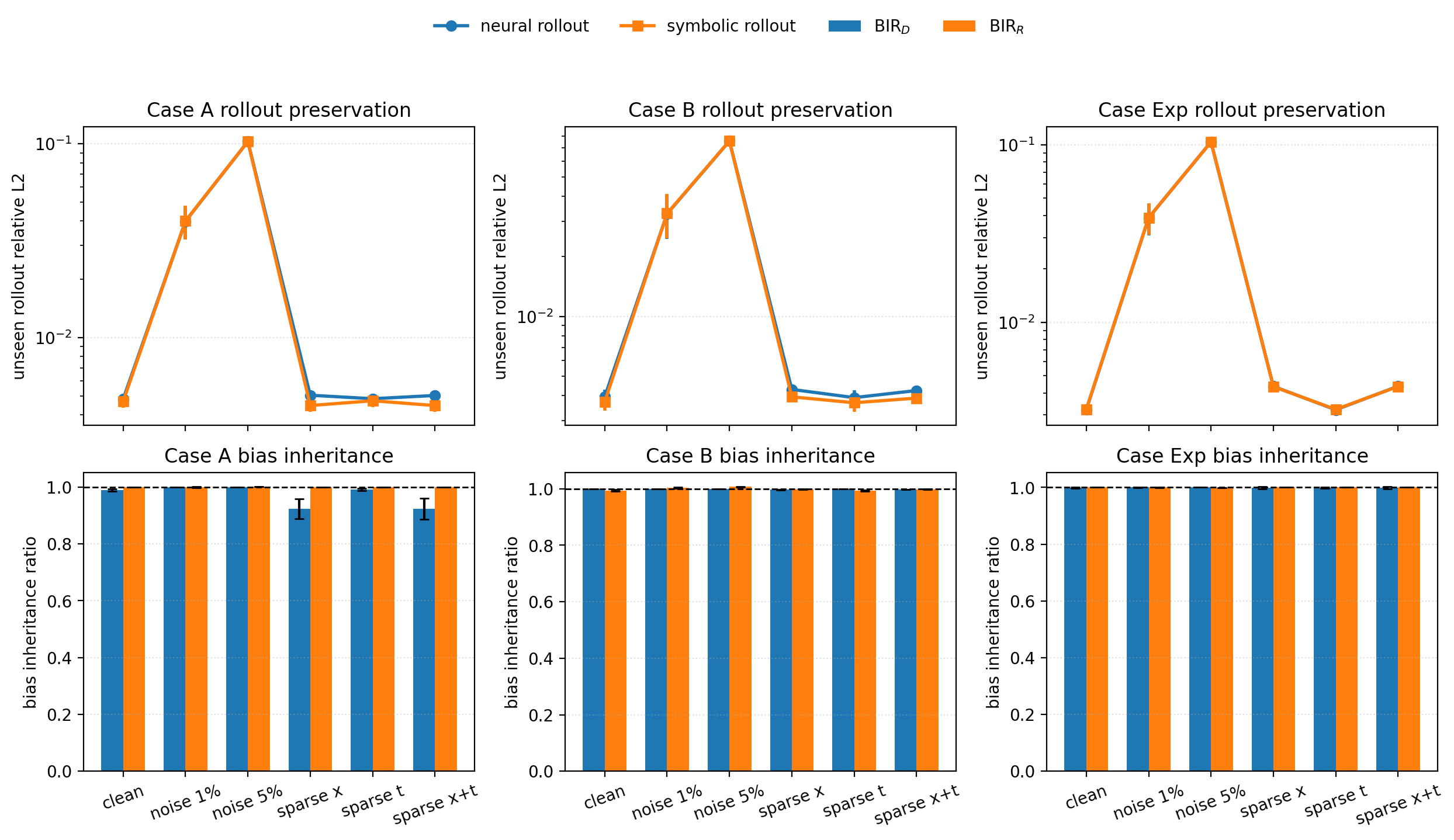}
\caption{Rollout preservation and bias inheritance. Top: symbolic unseen rollout closely follows neural unseen rollout. Bottom: the bias inheritance ratios remain near one, showing that symbolic compression preserves rather than repairs Stage~1 constitutive bias.}
\label{fig:rollout_bir}
\end{figure}

Figure~\ref{fig:rollout_bir} shows two key dynamical consequences of symbolic compression on the reported settings. First, the symbolic rollout remains very close to the neural rollout across clean, noisy, and sparse settings. Second, the bias inheritance ratios for both diffusion and reaction remain near one. Together these results support the claim that symbolic compression is not correcting the constitutive law; it is distilling the numerical surrogate into a more compact representation while preserving its existing bias.

\section{Discussion}

The results support a more careful view of neural-symbolic closure discovery than the usual ``neural beats symbolic'' narrative. In matched-library settings, weak polynomial baselines are genuinely strong. This is not a pathological edge case for the neural model; it is consistent with what one should expect when the true closure lies inside the baseline hypothesis class and the induced weak-form system is sufficiently informative.

The mismatch regime is where the neural stage becomes scientifically meaningful. Even there, however, the role of the neural model is not to guarantee exact law discovery. Its role is to provide a flexible numerical constitutive surrogate when the restricted prior library is misspecified. The surrogate is therefore an intermediate object: useful, but not yet a validated scientific law.

The Stage~1 objective itself should also be interpreted carefully. A fixed-dataset ablation on Case~Exp clean data (Appendix Table~\ref{tab:objective_ablation}) shows that a purely weak-form objective performs worst on reaction recovery, weak loss, and unseen rollout, while removing only the strong-form term yields the worst diffusion recovery. In other words, the implemented Stage~1 objective is not uniformly optimal across metrics, but neither is it reducible to weak-form training alone. This is why the paper describes Stage~1 as a weak-form-driven hybrid rather than as a pure weak-form learner.

This is why the symbolic stage must be interpreted carefully. The reported experiments show that symbolic compression is highly effective as a distillation step. It can reduce constitutive curves to compact symbolic forms with very small additional surrogate error and almost no extra rollout degradation on the tested settings. But once that compression error is small, it ceases to be the dominant source of error in those settings. The remaining constitutive bias is then dominated by mismatch and identification error from Stage~1. In this sense, the symbolic stage is not the main bottleneck in the current benchmarks; it is the place where the surrogate becomes interpretable, not the place where it becomes correct.

The noisy benchmarks sharpen this point rather than weaken it. Even with a weak-form-driven hybrid training signal, Stage~1 closure recovery deteriorates markedly under $5\%$ observation noise, especially for the reaction term. This does not contradict the paper's thesis; it reinforces it. Weak-form training mitigates spatial differentiation brittleness, but it does not eliminate the inverse-problem difficulty of constitutive recovery under noisy, weakly informative observations. More broadly, optimization and conditioning difficulties are well known in physics-informed inverse solvers \cite{Wang_2022}. Table~\ref{tab:pysr_check} makes the same point from the Stage~2 side: even when the restricted symbolic family is replaced by the much more expressive PySR engine, the noisy constitutive bias is still inherited rather than repaired on this mismatch stress test. Figure~\ref{fig:noise_sweep_all_cases} sharpens the Stage~1 diagnosis across all three 1D cases: relative to the clean reference, the mean diffusion errors jump from $(6.93, 10.0, 4.77)\times 10^{-2}$ to $(4.61, 4.80, 3.97)\times 10^{-1}$ at $1\%$ noise for Cases A, B, and Exp, while the reaction errors jump from $(3.13, 2.05, 4.40)\times 10^{-1}$ to $(1.06, 1.79, 2.27)\times 10^{0}$. Most of the remaining deterioration occurs by $2\%$--$3\%$ noise (nominal SNR about $34$--$30$ dB), after which the curves largely saturate. This is better described as a protocol-specific degradation knee than as a universal critical threshold. Additional diagnostic ablations indicate that this knee should not be attributed to the space-weak residual alone. The current Stage~1 pipeline contains multiple explicit time-derivative channels, including the auxiliary strong residual and mass-balance terms in addition to the space-weak estimate of $u_t$, and the strongest noise sensitivity appears to come from that broader hybrid design rather than from the weak residual in isolation. The central scientific point is therefore structural: once Stage~1 has already lost the correct constitutive signal in the $1\%$--$3\%$ regime, a stronger Stage~2 engine such as PySR still inherits that degraded law in our current checks rather than repairing it. That shifts the scientific burden toward more robust Stage~1 numerical constitutive recovery under noise, potentially including fuller space-time weak formulations or other designs that reduce explicit time-derivative estimation, rather than increasingly elaborate symbolic post-processing.

These observations suggest a clear agenda for future work. The main difficulty is reliable numerical constitutive recovery under limited excitation, sparse observations, and misspecified prior structure. Better Stage~1 recovery may come from richer excitation design, stronger identifiability diagnostics, adaptive trajectory selection, improved regularization, or more problem-aware surrogate classes. In a few matched-library sparse settings, the diffusion error improves slightly relative to the clean reference, which is consistent with a mild implicit smoothing effect rather than a genuine information gain. In the current mismatch experiments, another asymmetry emerges: spatial coarsening is more harmful than temporal coarsening, suggesting that state-support and gradient information are more sensitive to spatial under-resolution than to a modest reduction in saved snapshot cadence. Appendix Figure~\ref{fig:case_exp_observation_breakdown} makes that asymmetry explicit for Case~Exp: temporal stride $2$ alone stays very close to the clean reference, while spatial stride $2$ accounts for almost all of the sparse-setting degradation. Once the numerical recovery problem is better controlled, the symbolic stage can remain simple and restricted.

The symbolic library is also intentionally narrow. It is designed to answer a specific question: can a learned surrogate be compressed into a compact family without changing its constitutive or dynamical behavior? It is not intended as a complete search space for nonlocal, singular, or history-dependent closures. If broader closure classes are needed, they should be introduced as controlled extensions of the same compress-and-reinsert protocol rather than as unvalidated formula mining.

The current study is intentionally narrow. Its main benchmark suite focuses on 1D periodic reaction-diffusion systems, synthetic data, and a controlled set of closure families. This keeps the mechanisms visible. Appendix Table~\ref{tab:2d_bias_inheritance} adds a minimal 2D periodic dimensionality check, but it should be read as a preliminary stress test rather than a full higher-dimensional benchmark program. Extending the same logic to richer 2D/3D systems, transport-diffusion-reaction couplings, or more realistic observation models will introduce additional identifiability challenges, especially for anisotropic or multivariate closures, as is already familiar in data-driven closure modeling for fluid and continuum systems \cite{Duraisamy_2019,Brunton_2020}. That broader program is important future work, but it should be built on the present lesson: low residual is not enough, and symbolic compression should be judged by what it preserves.

\section{Conclusion}

In this study, we presented a weak-form-driven hybrid neural-symbolic framework for data-driven closure discovery in nonlinear reaction-diffusion systems characterized by known partial differential equation structures but unknown constitutive laws. This framework explicitly decouples the discovery process into numerical constitutive recovery, restricted symbolic compression, and robust forward validation via equation re-simulation. Contrary to the assumption that symbolic regression automatically repairs a biased neural surrogate, our theoretical and empirical analyses demonstrate that once the symbolic compression error is minimized, the resulting symbolic closure fundamentally inherits the constitutive bias already present in the learned numerical model. Consequently, the predictive dynamic rollout of the symbolic model closely mirrors that of the neural surrogate, yielding a bias inheritance ratio that remains persistently near unity across clean, noisy, and sparse observational environments. These consistent results indicate that the primary operational bottleneck in neural-symbolic closure discovery lies in the initial stage of numerical constitutive recovery, particularly under conditions of limited dynamic excitation and severe function-class mismatch. Although preliminary evaluations on a two-dimensional domain corroborate this underlying bias-inheritance mechanism, these initial tests serve as an exploratory stress test rather than a comprehensive high-dimensional validation. Therefore, future progress in data-driven computational mechanics must prioritize improving mathematical identifiability, establishing stronger validation protocols, and exploring broader closure families, all while ensuring that the subsequent symbolic compression stage remains restricted, scientifically interpretable, and rigorously validated through explicit forward dynamic simulation.

\bibliographystyle{elsarticle-num}
\bibliography{references}

\appendix

\section{Implementation Details}

\subsection{Synthetic cases and numerical solver}

All reported experiments use the 1D periodic solver in the accompanying codebase. Trajectories are generated by an RK4 time integrator applied to a conservative finite-volume-style discretization of
\begin{equation}
u_t = \partial_x \left( D(u) u_x \right) + R(u).
\end{equation}
The three main benchmark cases are
\begin{align}
\text{Case A:}\quad & D(u) = 0.01 + 0.05u, \qquad R(u) = u(1-u), \\
\text{Case B:}\quad & D(u) = 0.01 + 0.03u^2, \qquad R(u) = u - 1.5u^2 + 0.5u^3, \\
\text{Case Exp:}\quad & D(u) = 0.01 + 0.035(1-e^{-2.5u}), \nonumber \\
& R(u) = u(1.1e^{-1.4u} - 0.22).
\end{align}
Case~A and Case~B are matched to the low-order polynomial baseline library. Case~Exp is a smooth non-polynomial stress test for function-class mismatch.

\subsection{Training and observation protocol}

Unless otherwise stated, the baseline and symbolic benchmarks use $8$ training trajectories, random smooth Fourier initial conditions with $4$ modes, and the simulation grid
\begin{equation}
n_x = 64,\qquad \Delta t = 10^{-4},\qquad T = 0.1,
\end{equation}
with snapshots saved every $10$ time steps. The excitation comparison in Table~\ref{tab:excitation} uses a lighter protocol with $n_x=48$, $T=0.05$, snapshots saved every $5$ steps, and $6$ training trajectories. The cross-resolution benchmark in Table~\ref{tab:cross_resolution} uses the explicit fine and validation grids reported in its caption. The main surrogate is a two-branch MLP with hidden width $64$, depth $2$, and $150$ epochs of Adam optimization at learning rate $10^{-3}$.

Observation degradation is applied after simulation. A ``5\% noise'' setting means additive Gaussian noise with standard deviation $0.05\,\mathrm{std}(u)$, followed by clipping to the admissible state range. A ``sparse $x+t$'' setting means spatial stride $2$ and temporal stride $2$ in the observed trajectories. Figure~\ref{fig:noise_sweep_all_cases} further resolves the $0\%,1\%,2\%,3\%,4\%,5\%$ noise response across all three 1D cases. All tables report mean$\pm$standard deviation over three random seeds unless stated otherwise.

Unless otherwise stated, both polynomial baselines use diffusion degree $2$, reaction degree $3$, and ridge regularization strength $10^{-8}$. The weak polynomial baseline uses the same default test-function family as the paper-facing weak residual, namely $4$ Fourier modes plus $4$ bump functions.
For the paper hybrid Stage~1 objective, the default weights are $\alpha=0.1$, $\beta=1$, $\gamma=1$, $\eta=1$, and $\lambda=10^{-4}$. The main neural backbone uses $\tanh$ activations together with an explicit polynomial-residual parameterization: a degree-$2$ polynomial backbone for diffusion and a degree-$3$ polynomial backbone for reaction, each augmented by a small residual MLP. These settings are meant as a baseline reference rather than as a claim of exhaustive hyperparameter tuning. The public GitHub repository includes the training scripts, configuration dictionaries, and representative reproduction commands used to generate the paper tables; the main scientific point of the paper is the structural bias-inheritance mechanism, not extreme tuning of these specific weights.

\subsection{Stage-1 objective ablation}

Table~\ref{tab:objective_ablation} reports a small fixed-dataset ablation on Case~Exp under the clean protocol. The three compared objectives are: pure weak-form training, weak-form plus auxiliary terms but without the strong residual, and the full paper objective. The result is intentionally mixed. The full hybrid objective gives the best diffusion recovery, the smallest weak loss, and the best unseen rollout, while the weak-plus-auxiliary objective without the strong term gives the best reaction error. This is consistent with the main-text claim that the implemented Stage~1 learner is a weak-form-driven hybrid whose behavior is metric-dependent rather than uniformly dominant.

\begin{table}[h]
\centering
\small
\resizebox{\textwidth}{!}{%
\begin{tabular}{lcccc}
\toprule
Objective & ErrD & ErrR & Weak loss & Unseen \\
\midrule
weak only & $9.072\mathrm{e}{-02} \pm 2.879\mathrm{e}{-04}$ & $7.556\mathrm{e}{-01} \pm 2.361\mathrm{e}{-03}$ & $1.151\mathrm{e}{-03} \pm 1.077\mathrm{e}{-05}$ & $7.231\mathrm{e}{-03} \pm 3.229\mathrm{e}{-04}$ \\
weak + auxiliary, no strong & $1.021\mathrm{e}{-01} \pm 4.155\mathrm{e}{-04}$ & $2.480\mathrm{e}{-01} \pm 4.824\mathrm{e}{-03}$ & $7.673\mathrm{e}{-04} \pm 3.782\mathrm{e}{-06}$ & $5.637\mathrm{e}{-03} \pm 6.639\mathrm{e}{-05}$ \\
full paper hybrid & $4.859\mathrm{e}{-02} \pm 1.772\mathrm{e}{-04}$ & $4.369\mathrm{e}{-01} \pm 7.874\mathrm{e}{-03}$ & $3.241\mathrm{e}{-04} \pm 5.672\mathrm{e}{-06}$ & $3.319\mathrm{e}{-03} \pm 1.465\mathrm{e}{-04}$ \\
\bottomrule
\end{tabular}
}
\caption{Small Stage~1 objective ablation on a fixed clean Case~Exp dataset. Pure weak-form training is not competitive on this mismatch stress test. Removing the strong residual helps the reaction closure but still worsens diffusion recovery and rollout relative to the full paper objective.}
\label{tab:objective_ablation}
\end{table}

\subsection{Observation-degradation breakdown}

\begin{figure}[h]
\centering
\includegraphics[width=\textwidth]{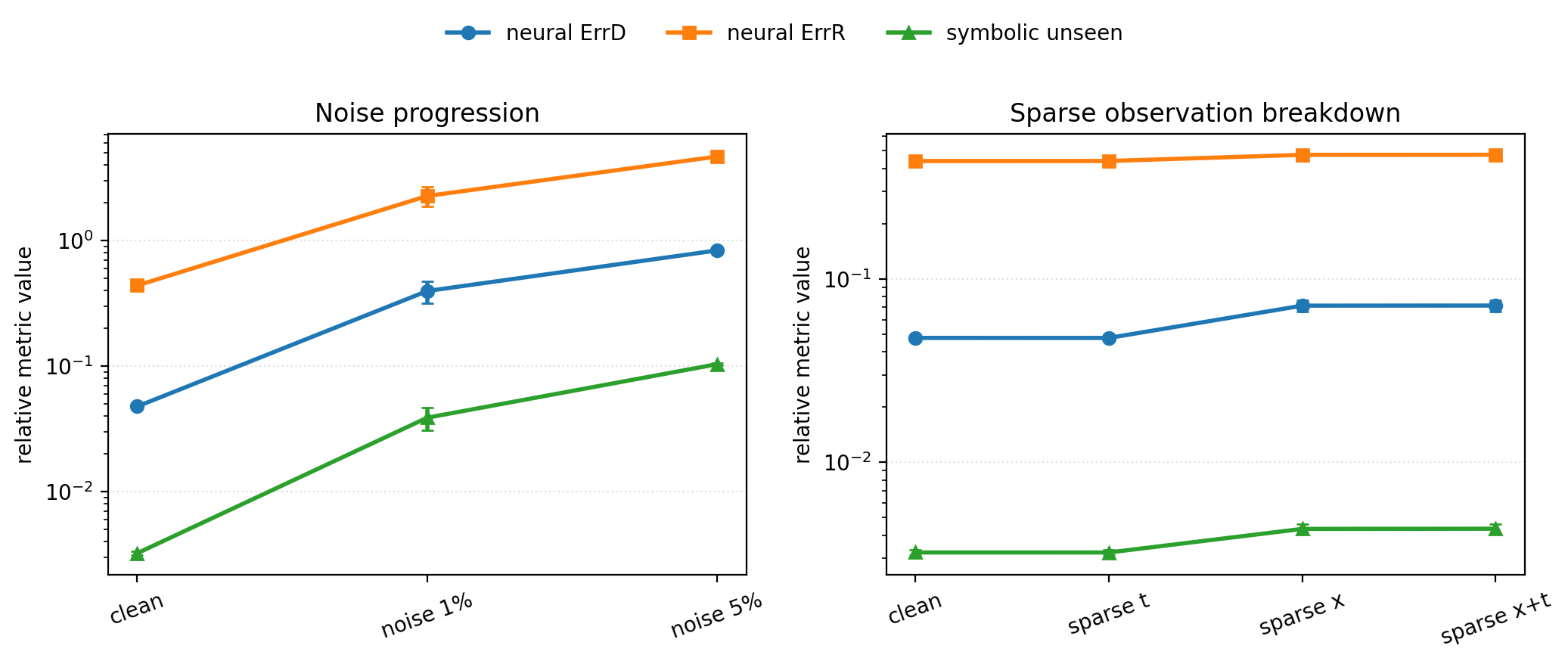}
\caption{Case~Exp observation-degradation breakdown. Left: representative clean, $1\%$, and $5\%$ noise checkpoints for the neural closure errors, together with the symbolic unseen rollout error. Right: temporal stride $2$ alone leaves the mismatch benchmark almost unchanged, while spatial stride $2$ accounts for nearly all of the sparse-setting degradation.}
\label{fig:case_exp_observation_breakdown}
\end{figure}

Figure~\ref{fig:case_exp_observation_breakdown} provides a complementary coarse view of the same Case~Exp degradation mechanisms. The left panel compresses the dedicated noise sweep into representative clean, $1\%$, and $5\%$ checkpoints while also showing the symbolic unseen rollout error. The right panel separates temporal from spatial downsampling: temporal stride $2$ alone changes little, whereas spatial stride $2$ drives nearly all of the sparse-setting penalty. This is consistent with the interpretation used in the main text, namely that the difficult part in this benchmark is preserving enough state-support and gradient information for Stage~1 recovery rather than merely preserving snapshot cadence.

\subsection{Minimal 2D dimensionality check}

Table~\ref{tab:2d_bias_inheritance} reports a deliberately small 2D periodic benchmark built from the same scalar closure families, but now with $32\times 32$ grids, $4$ training trajectories, and $120$ optimization epochs. The goal is not to claim a fully mature 2D suite. It is to test whether the core mechanism survives a first dimensionality increase. The results are consistent with that same mechanism. On both the matched Case~A and the mismatch Case~Exp, the symbolic stage again stays very close to the learned neural surrogate, with $\mathrm{BIR}_D$ and $\mathrm{BIR}_R$ remaining near one. On the noisy 2D Case~Exp stress test, Stage~1 deteriorates sharply, but Stage~2 still largely mirrors that degraded constitutive law rather than repairing it.

\begin{table}[h]
\centering
\small
\resizebox{\textwidth}{!}{%
\begin{tabular}{llcccccccc}
\toprule
Case & Setting & Neural ErrD & Neural ErrR & Symbolic ErrD & Symbolic ErrR & Neural unseen & Symbolic unseen & $\mathrm{BIR}_D$ & $\mathrm{BIR}_R$ \\
\midrule
Case A & clean & $8.293\mathrm{e}{-02} \pm 2.007\mathrm{e}{-03}$ & $3.188\mathrm{e}{-01} \pm 2.663\mathrm{e}{-02}$ & $8.084\mathrm{e}{-02} \pm 2.135\mathrm{e}{-03}$ & $3.188\mathrm{e}{-01} \pm 2.664\mathrm{e}{-02}$ & $2.614\mathrm{e}{-03} \pm 1.077\mathrm{e}{-04}$ & $2.007\mathrm{e}{-03} \pm 1.064\mathrm{e}{-04}$ & $9.748\mathrm{e}{-01} \pm 2.189\mathrm{e}{-03}$ & $9.999\mathrm{e}{-01} \pm 5.449\mathrm{e}{-05}$ \\
Case Exp & clean & $8.669\mathrm{e}{-02} \pm 1.458\mathrm{e}{-03}$ & $3.317\mathrm{e}{-01} \pm 3.214\mathrm{e}{-02}$ & $8.573\mathrm{e}{-02} \pm 1.524\mathrm{e}{-03}$ & $3.317\mathrm{e}{-01} \pm 3.214\mathrm{e}{-02}$ & $2.092\mathrm{e}{-03} \pm 8.974\mathrm{e}{-05}$ & $2.236\mathrm{e}{-03} \pm 8.949\mathrm{e}{-05}$ & $9.889\mathrm{e}{-01} \pm 9.944\mathrm{e}{-04}$ & $1.000\mathrm{e}{+00} \pm 6.712\mathrm{e}{-06}$ \\
Case Exp & noise 5\% & $5.719\mathrm{e}{-01} \pm 2.168\mathrm{e}{-02}$ & $3.783\mathrm{e}{+00} \pm 5.565\mathrm{e}{-02}$ & $5.719\mathrm{e}{-01} \pm 2.178\mathrm{e}{-02}$ & $3.780\mathrm{e}{+00} \pm 5.545\mathrm{e}{-02}$ & $2.437\mathrm{e}{-02} \pm 1.270\mathrm{e}{-03}$ & $2.440\mathrm{e}{-02} \pm 1.304\mathrm{e}{-03}$ & $1.000\mathrm{e}{+00} \pm 3.079\mathrm{e}{-04}$ & $9.994\mathrm{e}{-01} \pm 4.523\mathrm{e}{-05}$ \\
\bottomrule
\end{tabular}
}
\caption{Minimal 2D periodic dimensionality check with three seeds. The 2D benchmark is intentionally small and not tuned as aggressively as the 1D main suite, but it remains consistent with the same mechanism: symbolic compression stays close to the learned neural surrogate and inherits its constitutive bias rather than correcting it.}
\label{tab:2d_bias_inheritance}
\end{table}

\subsection{Restricted symbolic families}

The symbolic stage is intentionally restricted. For diffusion, the candidate library contains polynomial families up to degree $3$, a $(2,2)$ rational family, an exponential-decay family, and a saturation family. For reaction, the candidate library contains polynomial families up to degree $4$, a $(2,2)$ rational family, a $u\exp(\cdot)$ family, and a $u/(1+\cdot)$ saturation family. The selected expression is the lowest-complexity candidate whose fit mean-squared error lies within a $5\%$ relative tolerance and a $10^{-8}$ absolute tolerance of the best candidate. This library is not meant to be exhaustive; it is a controlled interpretable compression family.

\subsection{Reproducibility notes}

The main result tables are generated from CSV summaries written by the benchmark scripts in the accompanying codebase, including symbolic-compression, excitation, and cross-resolution runs. Fixed random seeds are used throughout. Exact bitwise determinism across hardware is not claimed because PyTorch deterministic kernels are not globally forced, but repeated runs in the current environment reproduce the aggregate conclusions and the saved benchmark CSVs.

\subsection{Code availability}

The implementation, manuscript source, selected paper-facing table summaries, and figure assets are available at \codeurl.

\end{document}